\documentclass[fleqn,usenatbib]{mnras}

\usepackage[T1]{fontenc}

\DeclareRobustCommand{\VAN}[3]{#2}
\let\VANthebibliography\thebibliography
\def\thebibliography{\DeclareRobustCommand{\VAN}[3]{##3}\VANthebibliography}

\usepackage{graphicx}	
\usepackage{amsmath}	
\usepackage{amssymb}	
\usepackage{verbatim}   
\numberwithin{equation}{section} 

\usepackage{newtxtext,newtxmath}

\newcommand{\struutup}{\rule{0ex}{3.2ex}}

\newcommand{\vsini}{$v$~$\!\sin i$~}
\newcommand{\kms}{km\,s$^{-1}$}
\newcommand{\cd}{d$^{-1}$}
\newcommand{\spd}{$^{\rm d}{\hspace{-1.2mm}.}$ }
\newcommand{\spdb}{^{\rm d}{\hspace{-1.2mm}.} }

\title[]{\textcolor{black}{Updated modelling and refined absolute parameters of the oscillating eclipsing binary AS~Eri}\thanks{Based on observations made with the {\it Mercator} telescope, operated by the Flemish Community at the Observatorio del Roque de los Muchachos of the Instituto de Astrof\'{\i}sica de Canarias, La Palma, Spain, and the {\it Alfred Jensch} telescope at the Th\"uringer Landessternwarte, Tautenburg, Germany.}}

\author[Lampens P. et al.]{
P. Lampens$^{1}$\thanks{E-mail: patricia.lampens$@$oma.be},D. Mkrtichian$^{2}$,
H. Lehmann$^{3}$,
K. Gunsriwiwat$^{4}$,
L. Vermeylen$^{1}$,
J. Matthews$^{5}$,
and R. Kuschnig$^{6}$
\\
$^{1}$Royal Observatory of Belgium, Ringlaan 3, Brussel 1180, Belgium\\
$^{2}$National Astronomical Research Institute of Thailand, 260 Moo 4, T. Donkaew, A. Maerim, Chiang Mai 50180, Thailand\\
$^{3}$Th\"uringer Landessternwarte, Tautenburg, Germany\\
$^{4}$Department of Physics and Materials Science, Faculty of Science, Chiang Mai University, Muang, Chiang Mai 50200, Thailand\\
$^{5}$Department of Physics and Astronomy, University of British Columbia, 6224 Agricultural Rd, Vancouver, BC V6T 1Z1, Canada\\
$^{6}$Institute of Physics, Karl-Franzens University of Graz, NAWI Graz, Universit\"atsplatz 5/II, 8010 Graz, Austria
}

\date{Accepted XXX. Received YYY; in original form ZZZ}

\pubyear{2021}

\begin{document}
\label{firstpage}
\pagerange{\pageref{firstpage}--\pageref{lastpage}}
\maketitle

\begin{abstract}

{We present a new study of the Algol-type eclipsing binary system AS~Eri based on the combination of the MOST and TESS light curves and a collection of very precise radial velocities obtained with the spectrographs HERMES operating at the \textit{Mercator} telescope, La Palma, and TCES operating at the \textit{Alfred Jensch} telescope, Tautenburg. The primary component is an A3~V-type pulsating, mass-accreting star. We fitted the light and velocity data with the package PHOEBE, and determined the best-fitting model adopting the configuration of a semi-detached system. The orbital period has been improved using a recent (O-C) analysis and the phase shift detected between both light curves to the value 2.6641496~$\pm$~0.0000001 days. The absence of any cyclic variation in the (O-C) residuals confirms the long-term stability of the orbital period. Furthermore, we show that the models derived for each light curve separately entail small differences, e.g. in the temperature parameter T$_{\rm eff,2}$. The high quality of the new solutions is illustrated by the residuals. We obtained the following absolute component parameters: L$_1$ = 14.125~L$_{\odot}$, M$_1$ = 2.014~M$_{\odot}$, R$_1$ = 1.733~R$_{\odot}$, log g$_1$ = 4.264, L$_2$ = 4.345~L$_{\odot}$, M$_2$ = 0.211~M$_{\odot}$, R$_2$ = 2.19~R$_{\odot}$, log g$_2$ = 3.078~ with T$_{\rm eff,2}$/T$_{\rm eff,1}$ = 0.662~$\pm$~0.002. Although the orbital period appears to be stable on the long term, we show that the light-curve shape is affected by a years-long modulation which is most probably due to the magnetic activity of the cool companion. }
\end{abstract}



\section{Introduction}\label{sec:intro}
Eclipsing systems are essential objects for our understanding of the properties of stars as well as stellar systems. Well detached, double-lined, eclipsing systems offer the advantage of model-independent fundamental parameters of their components which can be used as direct constraints in the search for relevant models of stellar structure and evolution across the HR-diagram \citep{2010A&ARv..18...67T}. An additional constraint may come from the equal age, equal composition requirement. 
Their modelling furthermore allows to derive stellar surface properties such as reflection, gravity brightening and limb darkening coefficients, tidal flattening (ellipsoidality), surface inclinations (spin-orbit alignment) and even mode identifications obtained from a dynamical screening of the pulsating surfaces during eclipses (using the spatial filtering or dynamical eclipse mapping methods, \textcolor{black}{e.g. hereunder}) \citep{2016ApJS..227...29P,2018ApJS..237...26H}. 

Eclipsing systems with pulsating components are most interesting study targets: not only do they provide the fundamental component properties needed in the search for a precise asteroseismic model, but they also undergo a series of phenomena that are intrinsically linked to the gravitational forces acting on the components, for example tidal effects and mass transfer stages. These phenomena can and will influence the stellar interiors and surfaces, including the pulsations \citep{2021Galax...9...28L}. 
The orbital configuration plays a role too since different kinds of tidal interactions are observed in short-period, circular versus eccentric systems. Tides can generate stellar pulsations, e.~g. by a resonance mechanism such as in the eccentric 'heartbeat' systems \citep[][]{2012MNRAS.420.3126F,2020ApJ...903..122C}. In turn, tidally excited non-radial oscillations can also affect the evolution of close binaries \citep[][Sect.IV.F and references therein]{2021RvMP...93a5001A}. 

\textcolor{black}{The first pulsating mass-accreting components of semi-detached Algol-type systems, AB~Cas \citep{1971IBVS..596....1T} and Y~Cam \citep{1973IBVS..823....1B}, were discovered in the 1970's. However, understanding that they belong to a new class of pulsating stars was delayed until the early 2000's, when \citet{2002ASPC..259...96M,2004A&A...419.1015M} classified them under the name 'oEA stars' as a special class of short-period main-sequence mass-accreting pulsators that are evolutionary different from the classical $\delta$ Scuti stars found in detached eclipsing systems.} \textcolor{black}{Since this recognition, the number of mass-accreting components discovered from the ground gradually increased, reaching {more than 70} members in 2018 \citep{2018MNRAS.475.4745M}. Recently, this number increased by a factor of about two thanks to the results of the TESS mission (Mkrtichian et al., in prep.)} \\

These close binary systems experience (still) on-going non-stationary mass transfer via the inner Lagrange L${1}$ point unto the atmosphere of the pulsating component. Mass accretion results in changes of the radius, mass, density as well as of the short-scale response of the pulsating star which depend on the mass-accretion rate. It is still a matter of debate whether or not the mass transfer stage in Algol-type systems is fully conservative or not \citep{2011MNRAS.418.1764B,2018A&A...615A.131L}.
\textcolor{black}{So far, RZ~Cas is the best investigated oEA system which makes it suitable for exploring the interaction between mass exchange and pulsations. \citet{2020A&A...644A.121L} detected two opposite, cool and dark spots on the surface of the secondary component facing the Lagrangian points L1 and L2 from their long-term spectroscopic study of RZ~Cas lasting from 2001 to 2017. They showed that the spot sizes varied in an opposite way with a characteristic time scale of 9 years (already reported from the (O-C) variations by \citet{2018MNRAS.475.4745M}), while the time scale of the L2 spot migration was found to be close to 18 years. They interpreted the 9-yr time scale as half of an 18-yr magnetic dynamo cycle of the cool companion.  
They also concluded that the mass-transfer rate is controlled by the variable depth of the Wilson depression in the magnetic spot around L1. These results illustrate the importance of an accurate determination of the fundamental parameters of components of eclipsing binary systems and their surface structures via precise photometric and spectroscopic analyses.}\\

AS~Eri (HD~57167, HIP~35487, HR~2788, TYC~5965-2336-1) is an 8th-magnitude, 
semi-detached eclipsing and double-lined spectroscopic binary 
\citep{1973ApJ...185..265P} of spectral type A3\,V\,+\,K0\,III with the light ephemeris given by \citet{2004AcA....54..207K,2005yCatp005005402K}~:\\ 
  \indent \indent \indent Min.~(HJD) = 2452502.108~+~E~$\times$~2\spd664145. \\
\citet{1960AJ.....65..139K} and 
\citet{1960MNSSA..19..150L} 
obtained the first photo-electric light curves of AS~Eri in the filters blue and yellow. Based on their data and \citet{1960AJ.....65..139K}'s photometric solution, 
 \citet{1971ApJ...166..373H} 
derived two possible models with the secondary component distorted close to its Roche limit. \citet{1973ApJ...185..265P} and 
\citet{1984A&A...141....1V} 
were the first ones to derive a full set of orbital parameters for the system.  \citet{2020arXiv200101292A} 
derived the absolute parameters by analysing the light curve from the Transiting Exoplanet Survey Satellite (TESS)\footnote{Since there is only scarce information on how the results presented in Table~1 were obtained and no radial velocities were included, we consider this light curve solution as a preliminary result.}. According to  \citet{2002ApJ...575..461E}, 
this system evolved to its present configuration after a substantial loss of angular momentum.

Rapid pulsations with a period of 24.4~min (f = 59.0312~\cd), attributed to its primary component, were discovered by 
\citet{2000IBVS.4837....1G}. Later, \citet{2004A&A...419.1015M} 
confirmed the 24.4~min period and reported multiperiodic oscillations with additional frequencies of 62.5631 and 61.6743~\cd. AS~Eri is \textcolor{black}{one of the first five} semi-detached eclipsing binary systems with $\delta$ Scuti-type pulsations assigned to the class of oEA stars. 
\citet{2004A&A...419.1015M} 
proposed that, given their high inclination, the oEA systems are prime candidates for the detection and analysis of non-radial high-degree sectoral pulsations as well as for applying 'spatial filtering' for mode identification during the eclipses \citep{2003ASPC..292..369G,2004MNRAS.347.1317R}. 
In particular, the orbital-to-pulsation period ratio, $P_{\rm orb}/P_{\rm puls}$, of $\sim$157 indicates that AS~Eri~A is particularly interesting for an in-depth analysis of its {oscillatory} behaviour. 

\textcolor{black}{\citet{2008AJ....136.1736G} measured the rotational velocities of 23 eclipsing and spectroscopic binary systems using the techniques of Least-Squares Deconvolution (LSD) and Fourier analysis of the line profiles. They obtained \vsini's of 36 $\pm$ 3 and 40 $\pm$ 3~\kms, respectively for the primary and the secondary component of AS~Eri.} \citet{2013PASJ...65..105N} 
carried out an abundance analysis of the primary component of AS~Eri based on spectral data obtained with the High Dispersion Spectrograph (R $\sim$ 72\,000) at the Subaru telescope. \textcolor{black}{He reported underabundances of -0.66, -0.60, -0.57, -0.48, and -0.31~dex in respectively Fe, Ca, Mg, Ti, and Cr.}\\

In this work, we will determine the best-fitting model for the system based on two sets of space-based light curves supplemented by a well-distributed series of recently acquired, high-resolution spectroscopic data. \textcolor{black}{A detailed pulsational analysis of AS~Eri based on the space-based data (MOST and TESS), multi-site ground-based photometry as well as complementary high-resolution spectroscopy will be presented in a follow-up paper.} \\ 

\section{Observations and data reduction}\label{sec:obs}

\subsection{The radial velocity data\label{sec:RVs}}

High-resolution spectra of AS~Eri were collected in the years 2011, 2014 and 2015 for phase-resolved radial velocity monitoring. The observations were performed with the high-resolution fibre-fed \'echelle spectrograph \textsc{HERMES} (High Efficiency and Resolution Mercator Echelle Spectrograph, 
\citet{2011A&A...526A..69R}) mounted at the focus of the \mbox{1.2-m} {\it Mercator} telescope located at the international observatory 
{\rm Roque de los Muchachos} (La Palma (LP), Spain). The instrument is operated by the University of Leuven under the supervision of the \textsc{HERMES} Consortium. It records the optical spectrum in the range $\lambda$ = 377\,- 900\,nm across 55 spectral orders in a single exposure. The resolving power in the high-resolution mode is R = 85\,000. Technical details and the performance of the instrument are described in \citep{2011A&A...526A..69R}. 
\textcolor{black}{We also used the TCES spectrograph\footnote{https://www.tls-tautenburg.de/TLS/index.php?id=31\&L=1} mounted at the Coud\'e focus of the 2-m {\it Alfred Jensch} telescope at the Th\"uringer Landessternwarte (TLS), Tautenburg. The instrument is a high-resolution \'echelle spectrograph and covers the wavelength range 445\,- 755\,nm with a resolving power R = 58\,000 in combination with the NBI camera. \\} 

\begin{figure}
\center
\includegraphics[width=8.2cm,clip=]{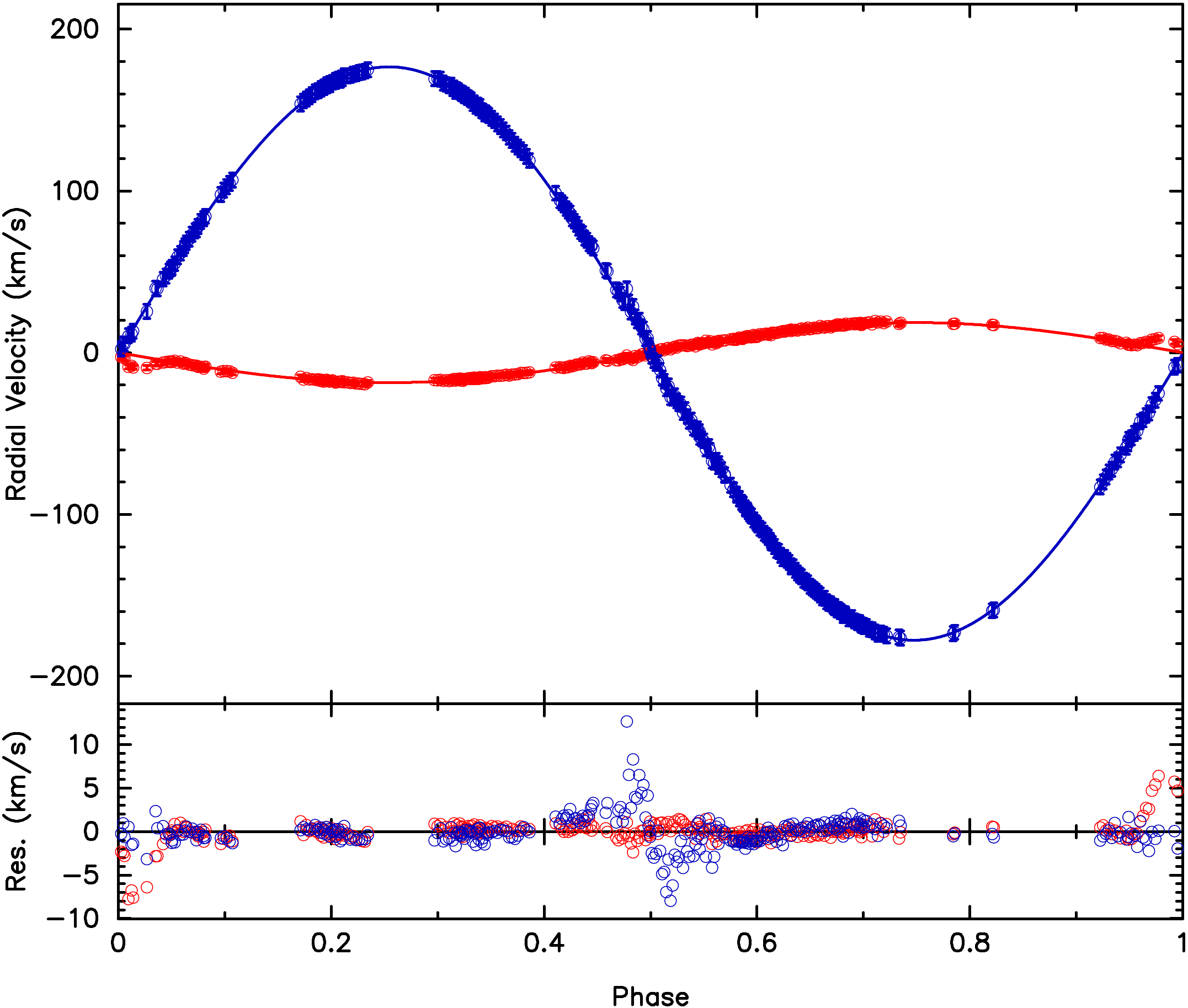} \\
\caption{Component radial velocities and pure RV-based orbital solution. The Rossiter-McLaughlin effect is clearly present in the residual data.}
\label{fig:ASEri:orbit0}
\end{figure}

In total, we acquired 164 {\sc HERMES} spectra in the years 2011, 2014 and 2015 whereas 166 {\sc TLS} spectra were collected in 2015. The radial velocities (RVs) were determined twice making use of 163 normalised {\sc HERMES} spectra and 160 normalised \textsc{TCES} spectra (7 spectra were omitted because of low S/N or artefacts), by one of us (HL) using his implementation of \textsc{TODCOR} which computes a two-dimensional (2D) cross-correlation \citep{1994Ap&SS.212..349M}, \textcolor{black}{as well as by using another implementation of \textsc{TODCOR} 
first applied in the study of AU~Mon (Desmet et al.\,2010)}. 
We also derived uncertainties on the component RVs of all the \textsc{HERMES} spectra using the latter implementation of \textsc{TODCOR}. 
A composite (A2+G2)-model spectrum with a (fixed) \vsini pair of (35, 40)~\kms~was used for reference. We next computed the standard deviations obtained from the RVs derived from 14 different spectral regions of width approx. 100 \AA~chosen within the interval 4150-5700 \AA~of the normalised {\sc HERMES} spectra. We stress the fact that the individual RVs derived in both cases are fully consistent with each other. {\textcolor{black}{Hereafter, we will use the RVs determined by HL.}
Table~\ref{tab:journal} lists the {barycentric} Julian dates of the available spectroscopic observations with the corresponding RVs {and their errors} for each component. 
A preliminary orbital solution based on \textcolor{black}{all the component RVs} collected with both spectrographs was computed
(see col.\,2 of Table~\ref{tab:orbPHE} and Fig.~\ref{fig:ASEri:orbit0}). Table~\ref{tab:orbPHE} (col.\,3) also shows the RV-based solution derived by \citet[][hereafter VH\&W]{1984A&A...141....1V} 
for comparison. Note that the Rossiter-McLaughlin (RML) effect was not modelled at this stage. }\\  

\begin{table}
\caption{Times (BJD) and radial velocities of the primary (RV1) and the secondary (RV2) components of AS~Eri, {with corresponding standard deviations or uncertainties (e$\_$RV1 and e$\_$RV2) (in \kms)}. {The full table is available online as Supplementary Material.} }
\begin{tabular}{cccccc}
\hline\noalign{\smallskip}
       BJD &	Site	& RV1&	e$\_$RV1 & RV2 & e$\_$RV2\\
\hline\noalign{\smallskip}
2455609.398200  & LP & -3.360&  1.455 & 165.429 & 2.000\\
2455609.402254  & LP & -3.649&  1.018 & 164.712 & 1.932\\
2455610.354684  & LP & 29.775&  1.402 &-154.801 & 2.044\\
2455610.358737  & LP & 29.263&  1.008 &-156.470 & 2.393\\
2455611.377055  & LP &  3.406&  1.25  &  95.017 & 2.5  \\
2455611.381108  & LP &  3.527&  0.515 &  96.513 & 1.659\\
2455612.347864  & LP &  5.732&  1.681 &  78.090 & 4.625\\
2455612.351918  & LP &  6.346&  1.221 &  76.779 & 1.578\\
2455613.350182  & LP & 29.445&  0.830 &-147.308 & 2.643\\
2455613.354236  & LP & 29.289&  1.674 &-146.929 & 3.470\\
\ldots\\
\noalign{\smallskip}\hline\noalign{\smallskip}
\end{tabular}
\label{tab:journal}
\end{table}

\subsection{The space-based data sets\label{sec:LCs}}

We used two sets of light curves that were obtained respectively by the satellite missions MOST \citep{2016SPIE.9904E..2BR} and TESS 
\citep{2003PASP..115.1023W}. The MOST data set consists of 3875 relative magnitudes obtained in the MOST passband acquired between 2013, Oct. 10 and 2013, Nov. 20 ((BJD - 2\,400\,000) 56575.512841250 -- 56616.517707390). {An exposure time of 1.51 s/frame was used and 41 frames were stacked onboard,
giving a total exposure time of 61.9 s.} \textcolor{black}{We converted the original data expressed in relative mag to normalised flux (with respect to maximum light {at quadrature})} and removed one outlier.\\

The TESS data {for AS~Eri (TESS 301407485)} comprise 15936 data points in TESS light obtained in a single run from 2018, Oct. 19 to 2018, Nov. 14 (\textcolor{black}{Sector~4, (BJD - 2\,400\,000) 58410.90617810 -- 58436.83853767)}. {The exposure time of the flux time series data was 2 min.} We used the SAP (Simple Aperture Photometry) flux and its error, normalised and detrended the data, and removed three outlying data points. The light curves were phased against the published orbital period of 2.664145~d for a first check, confirming that both phase-folded plots looked {normal}. \\

\textcolor{black}{An additional and more recent light curve was acquired by TESS during the run from 2020, Oct. 22 to 2020, Nov. 16 (Sector~31, (BJD - 2\,400\,000) 59144.51980161 -- 59169.94926016). We did not use this light curve comprising 16156 SAP fluxes for the determination of the binary model. Instead, we will use it to verify the final solution obtained from the former data sets to conclude on the long-term stability of the light curve in Sect.~\ref{sec:mod}}. \\   

\begin{figure*}
\center
\begin{tabular}{@{}cc@{}}
  \includegraphics[width=8.8cm,clip=]{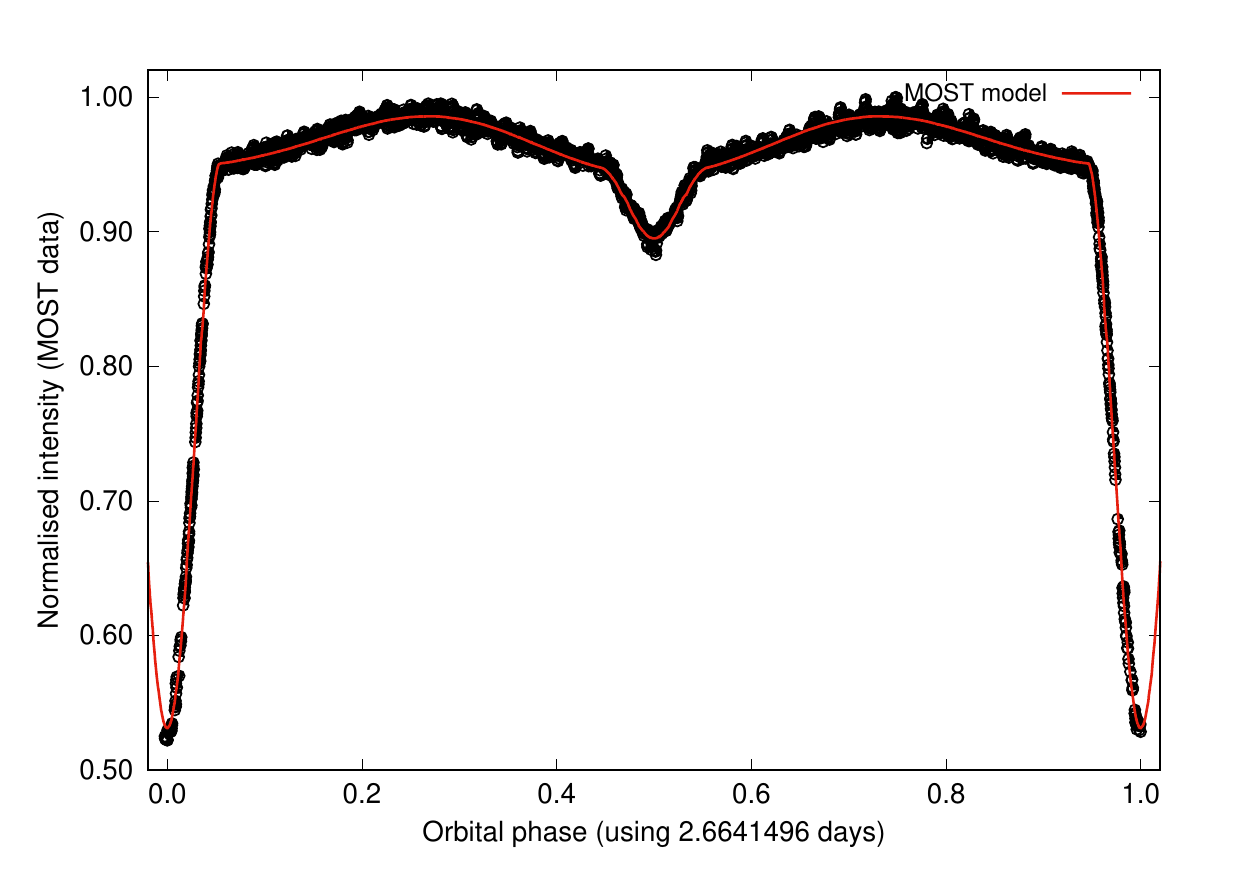} &
  \includegraphics[width=8.8cm,clip=]{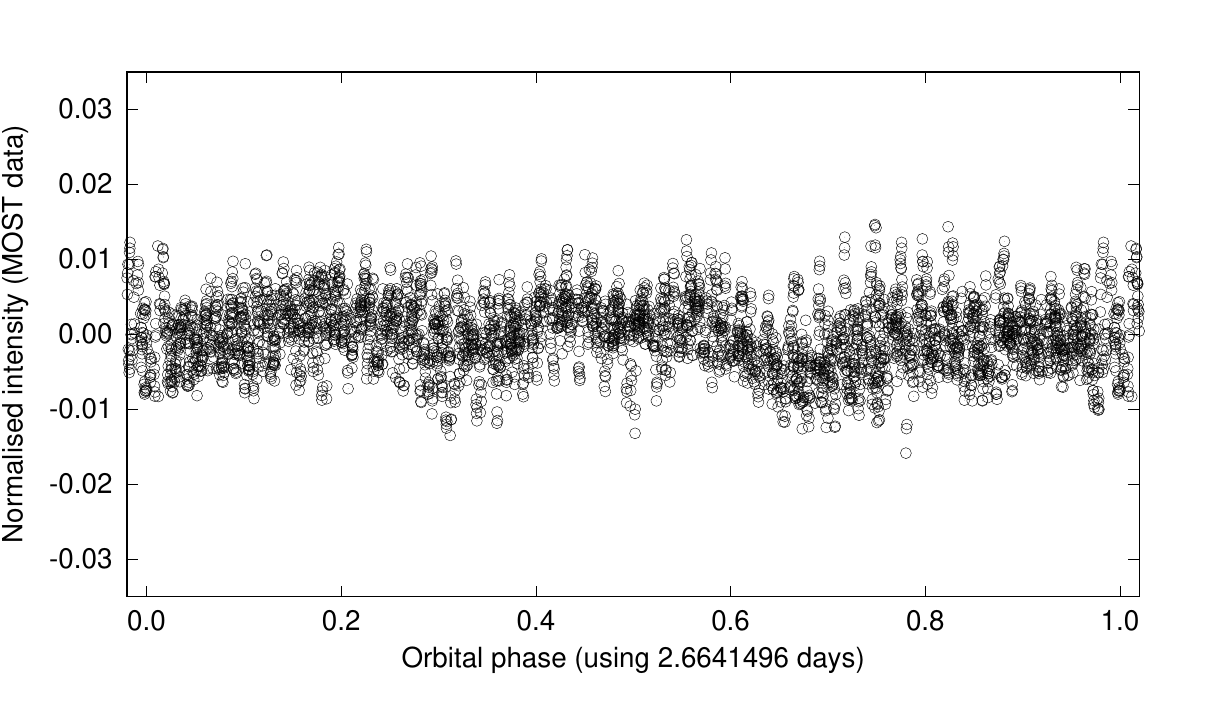} \\
  \multicolumn{2}{c}{}
\end{tabular}
\caption{Left: The MOST light curve and the corresponding model (solid red line) plotted against the orbital phase. Right: The residuals after model subtraction plotted against the orbital phase. }
\label{fig:ASEri:wMOST:LC1}
\end{figure*}

The component RVs were used as input for a simultaneous modelling study together with the \textsc{MOST} light curve. 
We used the eclipsing binary modelling software \textsc{PHOEBE} \citep{2011ascl.soft06002P}, 
initially without the estimated uncertainties, later with the standard deviations derived for the \textsc{HERMES} spectra complemented by mean uncertainties for the \textsc{TCES} spectra {estimated 
to be close to the typical (near average) values of the \textsc{HERMES} ones (we adopted 1.25 and 2.5~\kms~for components A and B, respectively).} For the \textsc{MOST} light curve, a mean standard error of 0.004 ($\sim$ 4~mmag) was derived from the scatter for flat portions of the light curve during the phases of light maximum and applied (see also Fig.~\ref{fig:ASEri:wMOST:LC1} where the residuals are displayed). During the first modelling experiments, the light ephemeris from 
\citet{2005yCatp005005402K} was adopted to compute the phases:
\begin{equation} 
\indent \indent \indent {\rm Min.(HJD)} = 2452502.108~+~{\rm E}~\times~2\spdb664145.
\end{equation} 
{At first look, we saw} no reason to suspect a change of the orbital period from the (O-C) diagram \textcolor{black}{published in the on-line \textsc{Atlas of (O-C) Diagrams of Eclipsing Binaries} (\textit{http://www.as.up.krakow.pl/ephem}}).

\section{Light and radial velocity modelling}\label{sec:bin} 

We used the eclipsing binary modelling software \textsc{PHOEBE-1.0} (legacy, dd. 08/07/2012, \textit{http://www.phoebe-project.org}). \textsc{PHOEBE-1.0} is based on the widely known Wilson-Devinney method \citep{1971ApJ...166..605W}. 
The code allows to simultaneously model the light and radial velocity curves of eclipsing binary systems using the full astrophysical information contained in the atmospheric models assuming a given configuration for the binary \citep{2005ApJ...628..426P}.
Here, we chose the semi-detached configuration with the secondary star filling its Roche lobe (MODE=5) and the preliminary orbital solution 
{(Table~\ref{tab:orbPHE}, col.\,2) as input for the search of a first combined model.} The adopted orbital period was initially 2.664145~days (see Sect.~\ref{sec:ominc}). 
For the relative \textcolor{black}{(passband-related) errors}, we adopted 1.0 for the \textsc{TODCOR} RVs {(using the initial standard deviations)} and 0.1 for the MOST and the TESS light curves  {(which increased the value of the cost function $\chi^{2}$ by a factor of 100, lending more weight to the photometric data sets).} The colours of the primary component have been evaluated by \citet{1973ApJ...185..265P}: 
he found (B - V) = +0.08. The primary is of spectral type A1~V or A3~V. We fixed the effective temperature of the primary component accordingly to the value of 8500~K. {We assumed standard values for the surface albedo's (1.0 and 0.5 for the primary and the secondary component, respectively), the gravity brightening of the primary component (1.0 as appropriate for stars with radiative envelopes as found by \citet{1924MNRAS..84..665V}
and generalized by \citet{1977A&A....58..267K})
and the limb darkening coefficients (logarithmic law) as obtained by interpolation in the tables by \citet{1993AJ....106.2096V}. }
The code provides the physical parameters of the components together with the orbital solution and formal errors from a combined, weighted least-squares analysis.  \\

\begin{table}
\begin{center}
\caption{Orbital elements with formal errors and physical parameters of AS~Eri (see Fig.~\ref{fig:ASEri:orbit0}). Col.~2) based on 323 newly acquired spectra. Col.~3) from the RV-solution derived by \citet{1984A&A...141....1V}.}
\begin{tabular}{@{}lcc@{}}
\hline\noalign{\smallskip}
   Orbital element   &  This work  &  VH\&W  \\ 
 \noalign{\smallskip}\hline\noalign{\smallskip}
 P (days)             &  2.6641534 $\pm$ 0.0000038 &  2.664152  \\
 T (periastron, 2400000.+)       &  55831.078 $\pm$ 0.079  & 28538.\textcolor{black}{076} \\
 e                    &  0.0107     $\pm$ 0.0021   &  null (fixed) \\
 i ($^{\circ}$)       &  undef.     &  80.451  \\
 $\omega$ (rad)        &  4.99    $\pm$ 0.19  & undef. \\ 
 V$_{01=02}$ (\kms)    &  11.622   $\pm$ 0.069  &  11.70 $\pm$ 0.17 \\
 q                     &  0.10470  $\pm$ 0.00067 & 0.1069 $\pm$ 0.0014 \\
 K$_1$ (\kms)            &  18.55   $\pm$ 0.11   &   ---  \\
 K$_2$ (\kms)            & 177.21   $\pm$ 0.42   &   ---  \\
 a$_{\rm 1}$.sin i (A.U.)   & 0.00454     $\pm$ 0.00003 & --- \\
 a$_{\rm 2}$.sin i (A.U.)   & 0.0434     $\pm$ 0.0001 & --- \\
 a.sin i (A.U.)             & 0.0479  & 0.0478  \\
 M$_{\rm 1}$.sin i$^3$ (${\rm M}_{\odot}$)  & 1.874    $\pm$ 0.013  &  ---   \\
 M$_{\rm 2}$.sin i$^3$ (${\rm M}_{\odot}$)  & 0.196    $\pm$ 0.002  &  ---   \\
\noalign{\smallskip}\hline\noalign{\smallskip}
 rms (\kms) (comp A \& B)  &   resp. 1.27 \& 1.80 &  ---  \\ 
\noalign{\smallskip}\hline\noalign{\smallskip}
 \end{tabular} 
 \label{tab:orbPHE}
 \end{center}
 \end{table}

\begin{table*}
\begin{center}
\caption{Orbital and physical parameters with formal errors, based resp. on the {\sc MOST} light curve and the component RVs, and on the same with the {\sc TESS} light curve added.} \struutup
\begin{tabular}{lcc}
\hline\noalign{\smallskip}
   Orbital element   &  {Phoebe-1.0} (w/o TESS) &  {Phoebe-1.0} (with TESS)\\
\noalign{\smallskip}\hline\noalign{\smallskip}
P (days)             &  2.664145 (fixed) & 2.664145 (fixed) \\
HJD$_{0}$ (2400000.+) &  52502.11297 $\pm$ 0.00007  & 52502.11613 $\pm$ 0.00002 \\
e                    &  0.00125  (free then fixed) & 0.00125 (fixed) \\
i ($^{\circ}$)       &  80.3250  $\pm$ 0.0056  &  80.4261 $\pm$ 0.0008 \\
$\omega$ ($^{\circ}$) & 4.696  $\pm$ 0.56  &  4.696 (fixed) \\
V$_{01=02}$ (\kms)    &  11.899 $\pm$ 0.028 &  11.899 (fixed) \\
q                     & 0.10486 $\pm$ 0.00013 & 0.10486 (fixed) \\
T$_{\rm eff,2}$ (K)   &   5609   $\pm$  8   &  5609 (fixed) \\
f$_{\rm 1}$           &   1.20  (fixed) & 1.20 (fixed) \\
g$_{}$b$_{\rm 2}$     &   0.077  $\pm$ 0.002 &  0.077 (fixed) \\
$\Omega_{\rm 1}$      &  6.1843  $\pm$ 0.0007 & 6.2251 $\pm$ 0.0003 \\
L$_{1,{\rm MOST}}$    &  10.1988 $\pm$ 0.0010 & 10.2028 $\pm$ 1.48 \\
L$_{1,{\rm TESS}}$    &    undef.    &  8.9955 $\pm$ 0.0003 \\
A (R$_{\odot}$)       & 10.5647  $\pm$ 0.0060 & 10.5647 (fixed)  \\
M$_1$ ($M_{\odot}$)  & 2.0176    &  2.0176   \\
M$_2$ ($M_{\odot}$)  & 0.2116    &  0.2116   \\
R$_1$ ($R_{\odot}$)     & 1.742   &  1.732   \\
R$_2$ ($R_{\odot}$)     & 2.198   &  2.198   \\
\noalign{\smallskip}\hline\noalign{\smallskip}
$\chi^2_{tot}$         &   1117 &  10.4e+09 \\ 
\noalign{\smallskip}\hline
\end{tabular}
\label{tab:PHEBE:fin}
\end{center}
\end{table*}

Table~\ref{tab:PHEBE:fin} presents the revised orbital elements and physical parameters with their formal errors associated to the best-fitting solutions obtained with \textsc{PHOEBE-1.0}, considering the given uncertainties and including two reflection effects. Columns 2 and 3 display the solutions respectively a) without and b) with the TESS light curve. Since the TESS (SAP) data are very numerous (N = 15542), they outnumber the {\sc MOST} data. The best-fitting solution which fits well all the data sets is thus heavily weighted toward the \textsc{TESS} light curve. In order to verify the stability of the orbital elements, a comparison with and without it was needed. We see that the final solution using the {\sc TESS} light curve is somewhat different from the solution which was based on the \textsc{MOST} light curve. This concerns, for example, the epoch (HJD$_{0}$) 
and the inclination (i) implying a small though significant change of some orbital parameters. Such a change might be caused by the presence of a third component (see VH\&W 
who concluded from their modelling on the possible existence of third light) or by apsidal motion (less probable in the case of a circular orbit). With respect to the phase shift (see the discrepancy in HJD$_{0}$), 
this may be easily explained by the need for an increase of order 4.5e-06~d in the orbital period. Such a correction is of the same order as the uncertainty on the period from each space light curve. A more accurate orbital period is difficult to assess from a multi-dimensional least-squares fitting because of the correlations between the free parameters. The result of further analysis (based on {adjusting} only one parameter) is described in Sect.~\ref{sec:ominc}.\\

\section{The revised orbital period}\label{sec:ominc} 

\begin{figure}
\center
 \includegraphics[width=9.0cm,clip=]{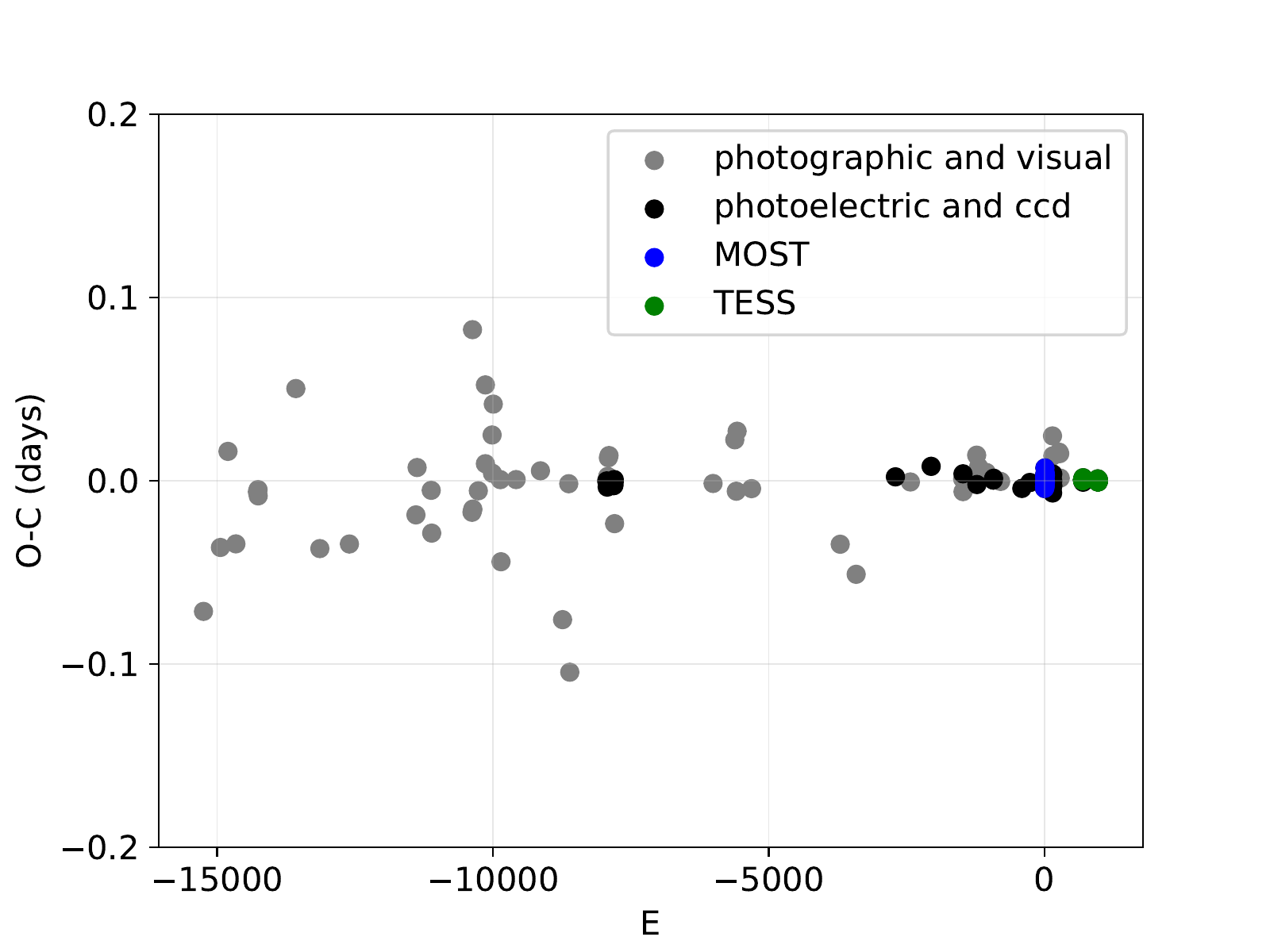} \\
\caption{Updated O-C diagram of AS~Eri. Different symbols illustrate the various types of times of minima. One obvious outlier ($>$ +0.3~d) was removed. } 
\label{fig:OminC}
\end{figure}

Before the final modelling work, we computed a revised light ephemeris based on all the times of minima collected with the photo-electric technique (E) and the CCD (C) \textcolor{black}{from the Lichtenknecker Database of the {\it Bundesdeutsche Arbeitsgemeinschaft f\"ur Ver\"anderliche Sterne e.V. (BAV)}}, together with new times of minima determined by us from the {\sc MOST} and {\sc TESS} light curves. 
\textcolor{black}{We furthermore included four unpublished times of minima obtained with the robotic telescopes \textsc{PROMPT-8} at Cerro Tololo Observatory, Chile, the \textsc{SSO-2} and \textsc{SSO-3} at Siding Springs Observatory, Australia, and the R-COP at Perth Observatory, New-Zealand.} {A detailed account of these ground-based observations will be presented in the follow-up paper devoted to the pulsational analysis of AS~Eri.} In this way, we gathered 104 times of primary and 30 times of secondary minima. Table~\ref{tab:OminC2}
lists all acquired times of minima. From a linear fit to these higher quality timings, we obtained the updated light ephemeris:
\begin{equation}
{\rm Min.(HJD)} = 2456575.58990~(\pm~3) + {\rm E} \times 2^{\rm d}{\hspace{-1.2mm}.}66415018~(\pm~3)
\label{eq:min}
\end{equation}


\noindent and adopted this new period to (re)compute the phases. \textcolor{black}{The corresponding diagram of (O-C) residuals (Table~\ref{tab:OminC2}, col.4), including all the data points except one outlier, is presented in Fig.~\ref{fig:OminC}.} 

Next, by fitting only the initial epoch HJD$_{0}$ of each light curve, we derived an accurate epoch of phase 0 for each data set. Since the difference between both initial epochs equals 1835.599061~d, the total phase shift of -0.00041~d and a refined period of 2.6641496 $\pm$ 0.0000001~d are obtained adopting the closest integer number of cycles (E = 689). This revised orbital period removes the previously reported phase discrepancy between both space light curves and was subsequently adopted during the final minimization run. \\

\begin{figure*}
\begin{tabular}{@{}cc@{}}
  \includegraphics[width=8.8cm,clip=]{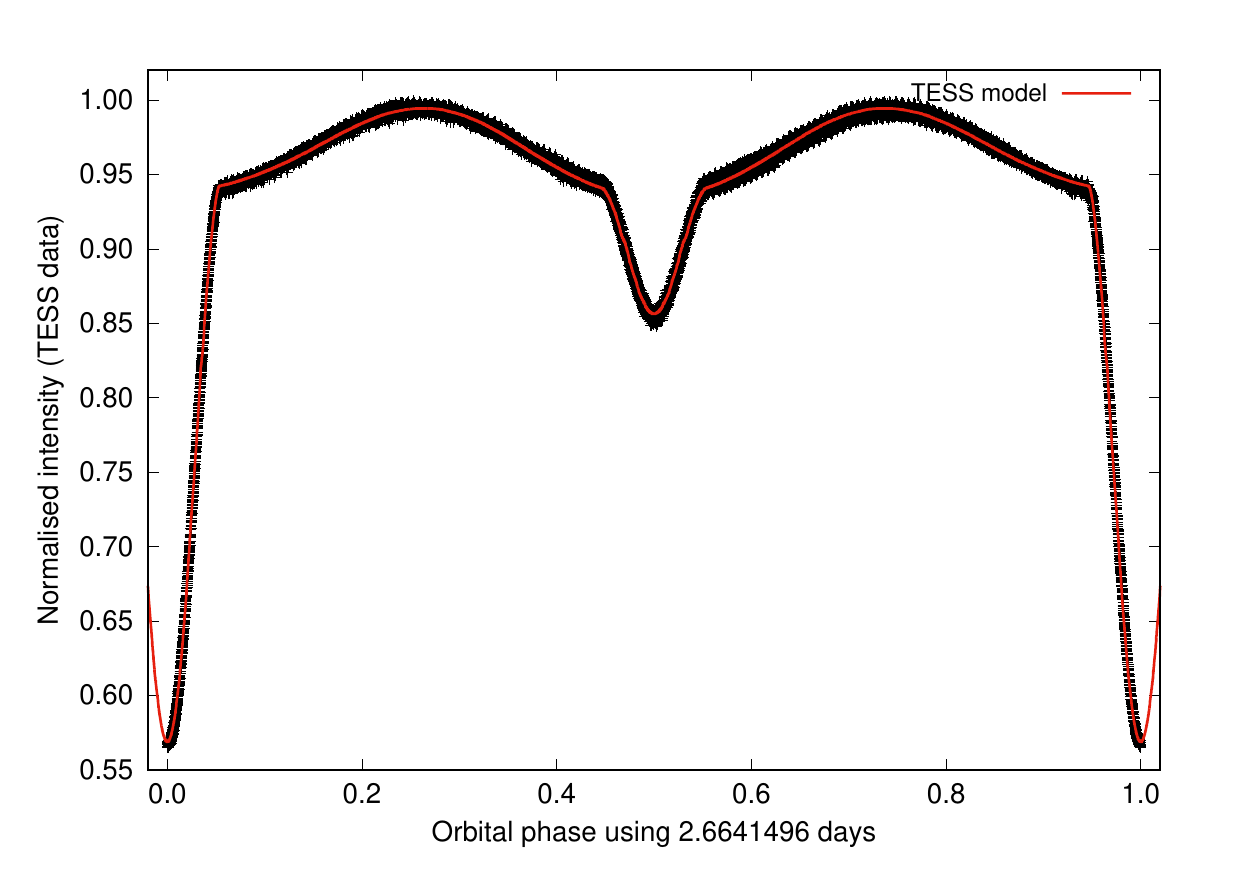} &
  \includegraphics[width=8.8cm,clip=]{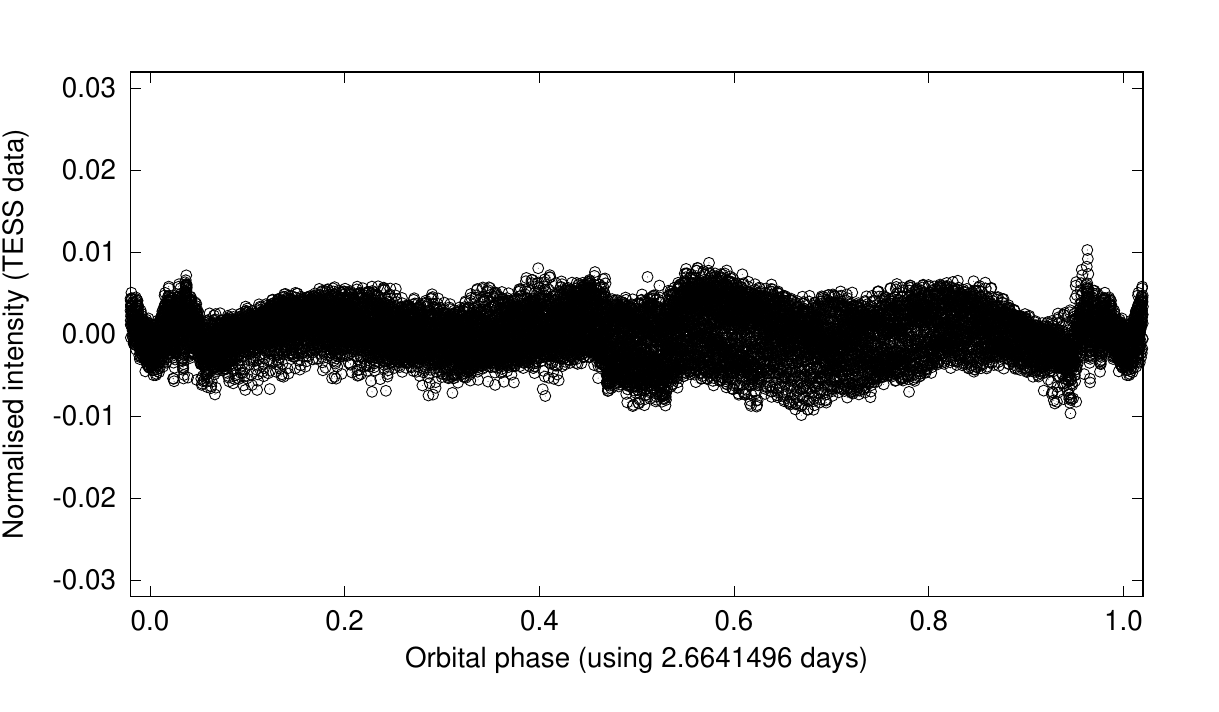} \\
  \multicolumn{2}{c}{}
\end{tabular}
\caption{Left: The detrended TESS light curve and its corresponding model (solid red line) plotted against the orbital phase. Right: The residuals after model subtraction plotted against the orbital phase. }
\label{fig:ASEri:wTESS:LC1}
\end{figure*}

\section{Light and velocity modelling: the final run}\label{sec:sim} 

For the final modelling experiment, we adopted the orbital period of 2.6641496~d (enabling to bridge the {phase} gap between both light curves) and we fixed the longitude of periastron to 4.70~rad (a value taken from the previous solutions, considering that it has little effect since the eccentricity is very small) and the filling factor f$_{1}$ to 1.1 - 1.2 (this provides an excellent match between the RV data and the model for the amplitude of the RML effect of comp~A, see Fig.~\ref{fig:ASEri:wTESS:resRV3}). We performed a new search for the best-fitting model twice: first, with the MOST light curve and the component RVs and then, with both (MOST+TESS) light curves and the component RVs.
\textcolor{black}{Here, we adopted the relative errors of 0.1 and 1.0 for the light curves and the \textsc{TODCOR} RVs, respectively.} \textcolor{black}{The logarithmic limb darkening law and two reflection effects were included in all our computations.}  All remaining parameters (except for T$_{{\rm eff},1}$) were set free at one point, though we used a stepwise approach to avoid the presence of strong correlations during the minimization process. 

Table~\ref{tab:PHEBE:fin2} presents the orbital and physical parameters with their formal errors associated to the best-fitting solutions obtained from a modelling with {\sc PHOEBE-1.0}. We list the final solution a) without (col.~2) and b) with the TESS light curve (col.~3). 
Figure~\ref{fig:ASEri:wMOST:LC1} illustrates the best-fitting binary model and the quality of the fit (illustrated by a plot of the residuals) based on the MOST light curve. 
Figure~\ref{fig:ASEri:wTESS:LC1} illustrates the best-fitting binary model and the quality of the fit (illustrated by a plot of the residuals) based on the TESS light curve. The residuals are in very good agreement with the estimated standard deviations of the respective data sets and the possible short-term drifts. The TESS light residuals are somewhat noisier than expected, because detrending of the data was performed linearly on the long-term time scale (small short-term drifts may still occur) and also \textcolor{black}{because of the existence of $\delta$ Scuti-type pulsations (see Sect.~\ref{sec:intro})}. \\

\begin{table*}
\begin{center}
\caption{Orbital and physical parameters with their formal errors from {\it Phoebe-1.0} based on the {\sc MOST} light curve with the RVs (col. 2), and on the same but with the {\sc TESS} light curve added (col. 3).} \struutup
\begin{tabular}{lcc}
\hline\noalign{\smallskip}
   Orbital element   &  {Phoebe-1.0} (w/o TESS) &  {Phoebe-1.0} (with TESS)\\
\noalign{\smallskip}\hline\noalign{\smallskip}
P (days)             &  2.6641496 (fixed) & 2.6641496 (fixed) \\
HJD$_{0}$ (2400000.+)  &  56575.58972 $\pm$ 0.00007  & 58411.18970 $\pm$ 0.00002 \\
e                    &  0.00125  (free then fixed) & 0.00063 $\pm$ 0.00008 \\
i ($^{\circ}$)       &  80.3246  $\pm$ 0.0063  &  80.4287 $\pm$ 0.0012 \\
$\omega$ (rad) & 4.70  $\pm$ 0.56  &  4.70 (fixed) \\
V$_{01=02}$ (\kms)    &  11.921 $\pm$ 0.028 &  11.918 $\pm$ 0.034 \\
q                     & 0.10468 $\pm$ 0.00032 & 0.10467 $\pm$ 0.00030 \\
T$_{\rm eff,2}$ (K)   &   5482   $\pm$  26   &  5646 $\pm$ 7 \\
f$_{\rm 1}$           &   1.20  (fixed) & 1.10 (fixed) \\
g$_{}$b$_{\rm 2}$     &   0.077  $\pm$ 0.002  &  0.075 $\pm$ 0.001 \\
$\Omega_{\rm 1}$      &  6.2479  $\pm$ 0.0011 & 6.2131 $\pm$ 0.0003 \\
L$_{1,{\rm MOST}}$    &  10.1953 $\pm$ 0.0013 & 9.0024 $\pm$ 1.44 \\
L$_{1,{\rm TESS}}$    &         undef.    &  8.9937 $\pm$ 0.0004 \\
A (R$_{\odot}$)       & 10.5535  $\pm$ 0.0059 & 10.5509 $\pm$ 0.0072 \\
M$_1$ ($M_{\odot}$)  & 2.0115    &  2.0124   \\
M$_2$ ($M_{\odot}$)  & 0.2106    &  0.2106   \\
R$_1$ ($R_{\odot}$)     & 1.722   &  1.729   \\
R$_2$ ($R_{\odot}$)     & 2.195   &  2.195   \\
\noalign{\smallskip}\hline\noalign{\smallskip}
$\chi^2_{tot}$         &   1085 &  10.3e+07 \\ 
\noalign{\smallskip}\hline
\end{tabular}
\label{tab:PHEBE:fin2}
\end{center}
\end{table*}

\section{The models}\label{sec:mod}

\begin{figure}
\center
 \includegraphics[width=8.8cm,clip=]{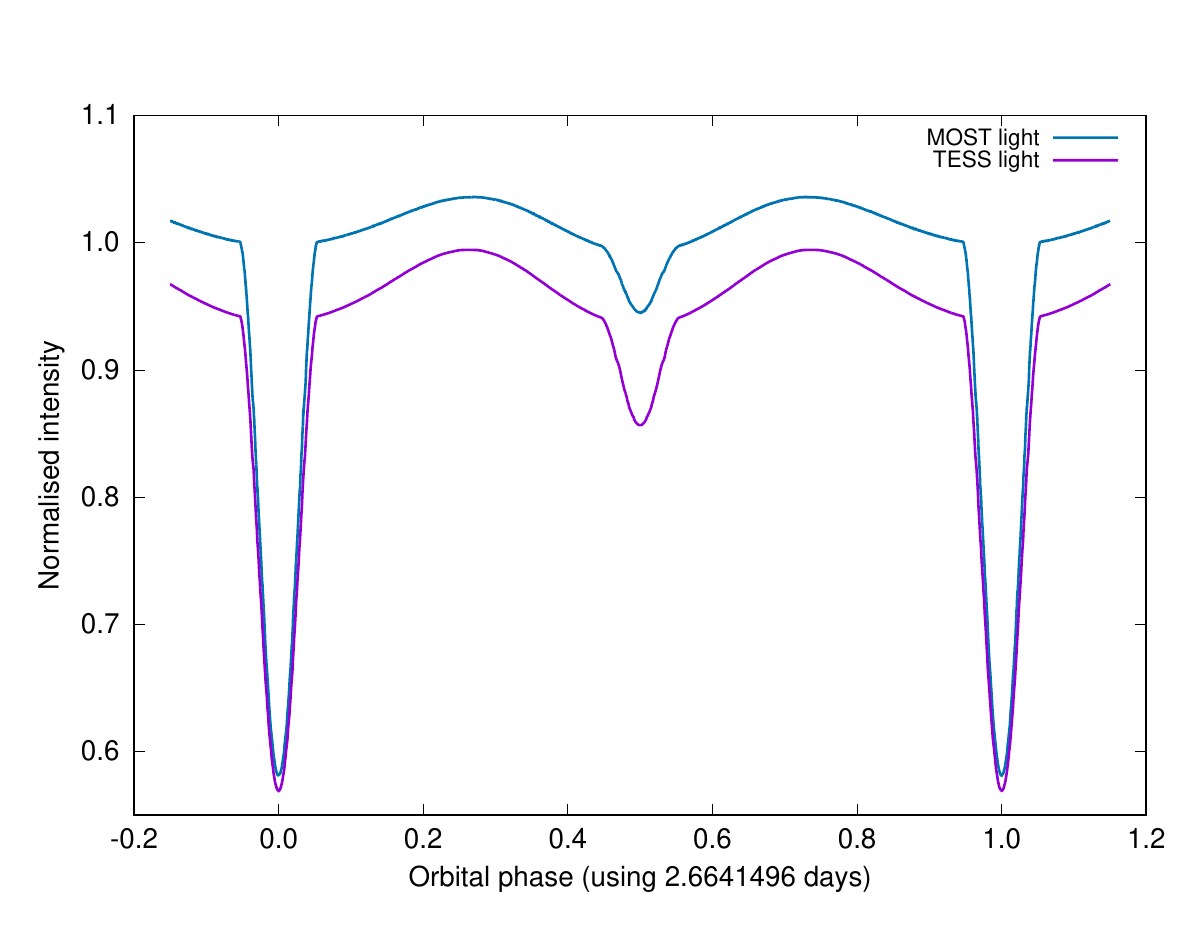} \\
\caption{Light curve models for AS~Eri. }
\label{fig:ASEri:LCmodel}
\end{figure}

\begin{figure*}
\center
\begin{tabular}{@{}cc@{}}
 \includegraphics[width=8.8cm,clip=]{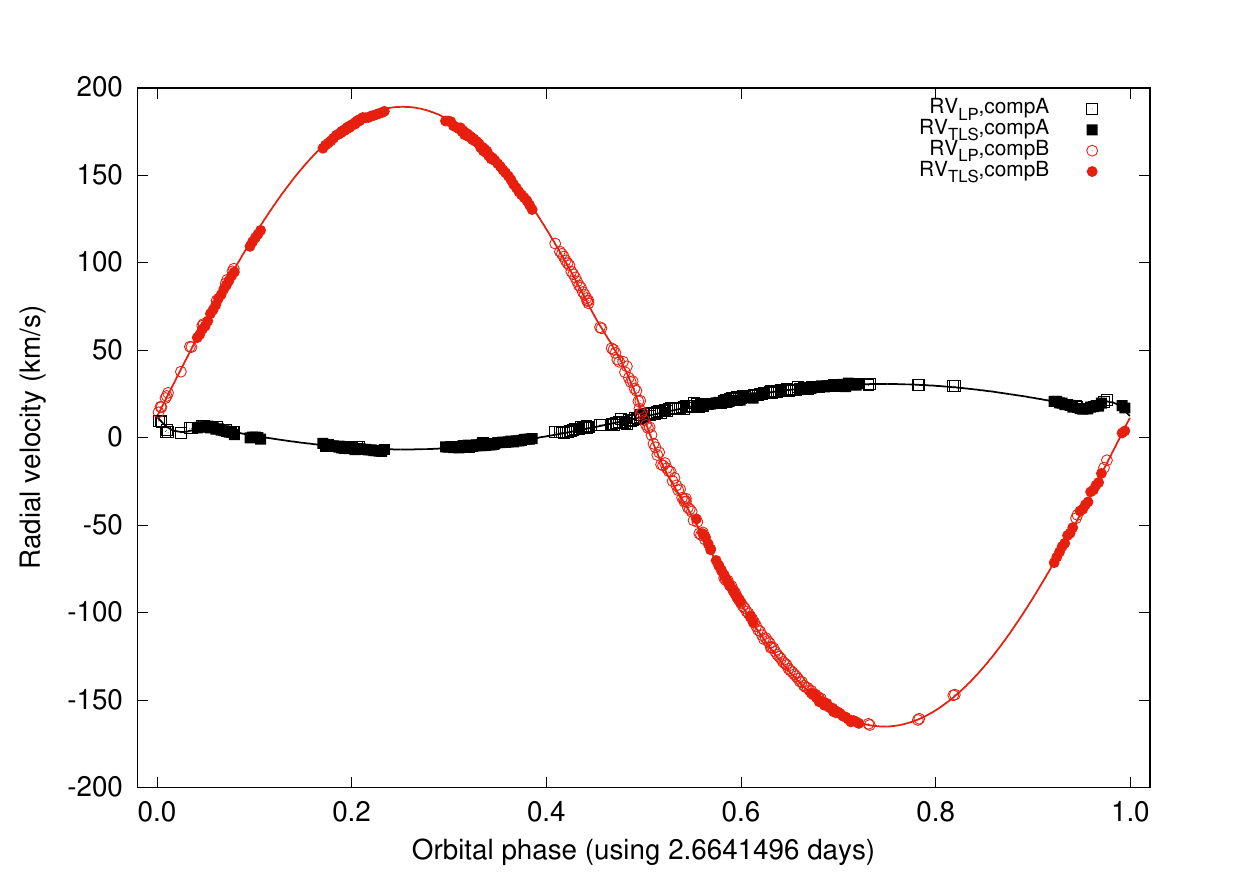} &
 \includegraphics[width=8.8cm,clip=]{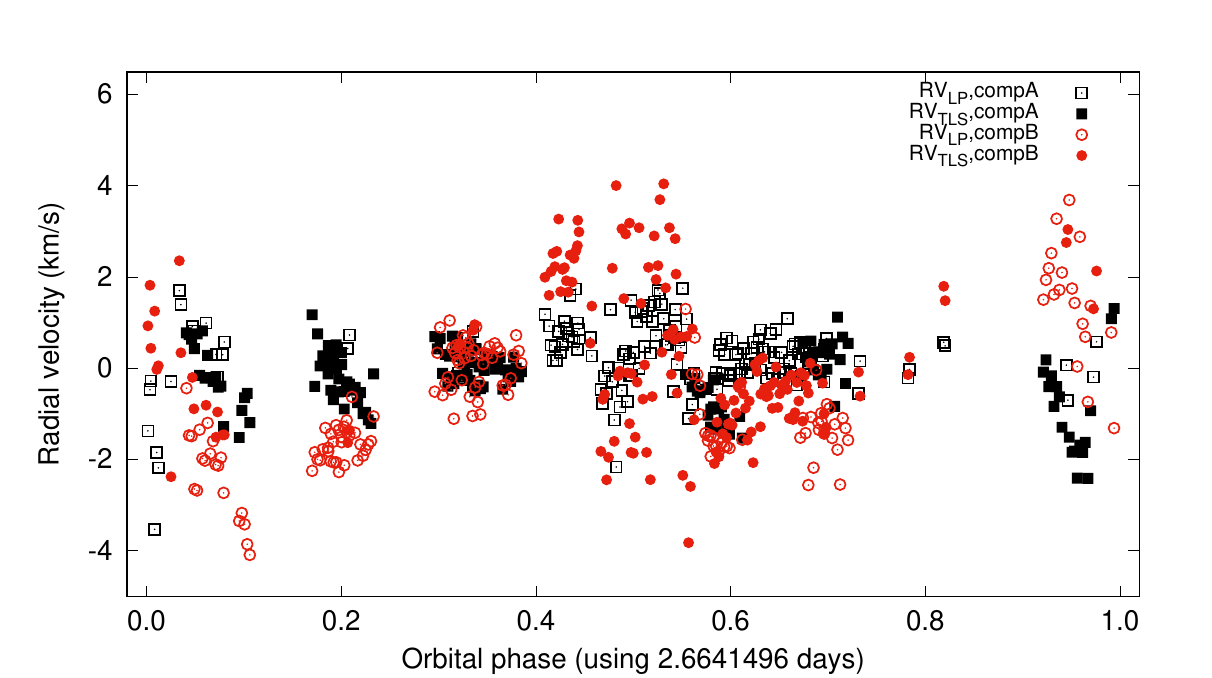} \\
 \multicolumn{2}{c}{}
\end{tabular}
\caption{Data and RV model (black and red lines) of AS~Eri. Right: The residuals after the model subtraction plotted against the orbital phase. {Unfilled symbols illustrate the HERMES RVs while filled symbols illustrate the TCES RVs.} } 
\label{fig:ASEri:wTESS:RV3+RVA3}
\end{figure*}

\begin{figure}
\center
 \includegraphics[width=8.8cm,clip=]{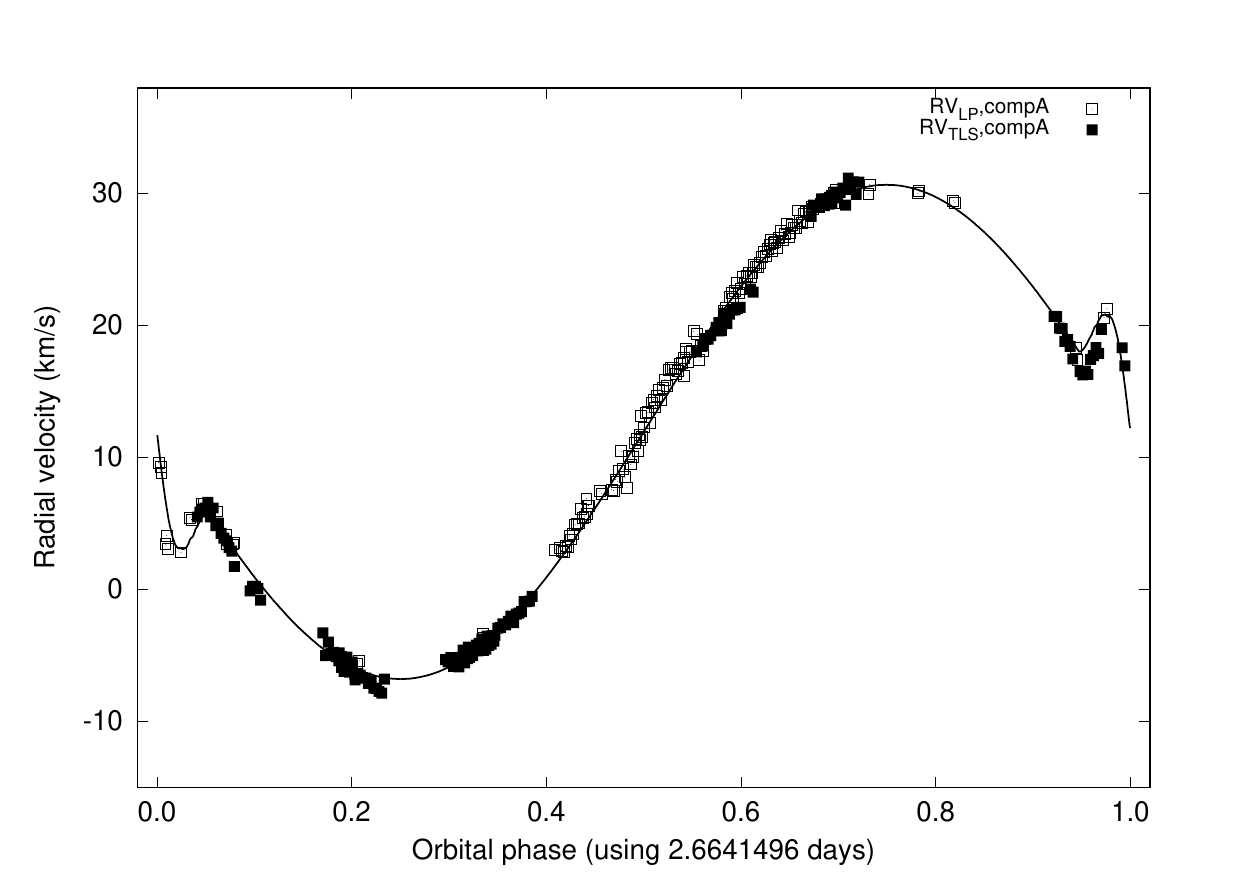} \\
\caption{Data and RV model (black line) for the primary component AS~Eri~A (zoom of Fig.~\ref{fig:ASEri:wTESS:RV3+RVA3} that illustrates the fit at the RML phase). {Unfilled symbols illustrate the HERMES RVs while filled symbols illustrate the TCES RVs.} }
\label{fig:ASEri:wTESS:resRV3}
\end{figure}

We present the solution associated to each light curve (LC), i.e. the models for the MOST and the TESS data sets together in Fig.~\ref{fig:ASEri:LCmodel}. It can be seen from Fig.~\ref{fig:ASEri:LCmodel} that a different LC model from each minimization was obtained. The difference between the LC solutions is not only due to the choice of the passband. \textcolor{black}{This is also evidenced by the small though significant changes in the values of the inclination (i), the gravitational potential ($\Omega_{1}$), and the effective temperature of comp~B (T$_{{\rm eff},2}$) in Table~\ref{tab:PHEBE:fin2}. The difference of 200~K in the temperature of comp~B does not seem large in se, but the LC models are clearly distinct. However, in both solutions, the same RV model was used.} The RV data and model are shown for both components in Fig.~\ref{fig:ASEri:wTESS:RV3+RVA3} (left panel), whereas Fig.~\ref{fig:ASEri:wTESS:resRV3} shows the same in a closer look for comp~A only. 
{The RV variations arise from the orbital motion and the changes of the visible surfaces of the components due to the aspherical shapes, the surface intensity distributions as well as the eclipses when the RML effect also occurs. The stellar shape is defined by the surface potential. \textsc{PHOEBE-1.0} uses the generalized surface potential considering elliptical orbits and asynchronous rotation \citep{1959cbs..book.....K,1979ApJ...234.1054W}. 
Besides the mass ratio, q, this potential also includes the fractional instantaneous separation between the two stars (to account for the eccentricity, e) and the synchronicity parameter f, i.e. the ratio between the rotational and the orbital angular velocity (to account for possible asynchronous rotation of the components) (cf. Eq.~3.31 in the book by \citet{2018maeb.book.....P}). 
In this case, the inclusion of asynchronous rotation provides an excellent fit for the RML effect of comp~A. }
The residual data for both components are also presented in Fig.~\ref{fig:ASEri:wTESS:RV3+RVA3} (right panel). \textcolor{black}{In the latter, we can see that the RV residuals of comp~B display small systematic shifts with respect to the final solution, e.g. at the orbital phases $\sim$ 0.1, in the vicinity of 0.5 and $\sim$ 0.9. In order to represent the component RVs with the latter solution in the finest possible details, we had to introduce a small eccentricity of 0.01063, i.e. almost the value of Table~\ref{tab:orbPHE}. This small {non-zero} eccentricity allowed us to significantly reduce the systematics in the residuals of the RVs of comp~B (at the phases $\sim$ 0.1 and 0.9), but it also affects the LC model such that the quality of the fit to the {(MOST+TESS)} data is no longer as good as before ($\chi_{tot}^2$ = 12.5e+07)}. Thus the final {combined} solution defines an almost circular orbit, whereas the fit to the RV curve of comp~B can be improved adopting a small eccentricity such as the one derived from a pure RV-based orbit. {Inspection of the corresponding RV residuals showed that such improvement may be purely formal and cannot be distinguished from Algol-related effects such as gas flows in the system and/or an inhomogeneous distribution of the circumbinary matter, thereby generating a false eccentricity in the orbital solution \citep[see also][]{1970PASP...82..815W}. }  
We conclude that the near-zero eccentricity found during the final modelling run due to the dominance of the {very} high-quality {(MOST+TESS)} light curves over the RV data {reflects the true system configuration}. \\

\textcolor{black}{In Table~\ref{tab:fund}, we provide the fundamental parameters and the relative shape (or radii) of each component {derived} from the final solution
{(col.~3 in Table~\ref{tab:PHEBE:fin2})}. The physical properties of each component were determined using all the data sets, including the TESS light curve from Sector~4, fitted together with an orbital period of 2.6641496~d. {The uncertainties were computed from various equivalent solutions obtained using (slightly) different starting parameter values.} } \\

\textcolor{black}{Finally, since the conclusion is that the LC model changed over the course of 5 years (i.e. the time span between the missions MOST and TESS), we decided to verify our solution based on the TESS light curve from Sector~4 against the more recent one from Sector~31. Figure~\ref{fig:ASEri:TESS:S31} illustrates this comparison. We observe that primary minimum is deeper again and that the more recent light curve is no longer agreeing with the TESS model. Its shape follows more closely that of the MOST model. From this, we deduce that the light-curve shape varies intrinsically over a time span of a few years. } \\

\begin{table}
\centering
\caption{Stellar parameters based on the final binary model. The relative radii expressed in units of the semi-major axis A are listed as 'radius'. }
\label{tab:fund}
\begin{tabular}{@{}lccr@{}}
\hline\noalign{\smallskip}
  Mass (M$_{\odot}$) &  Radius (R$_{\odot}$) &  M$_{\rm Bol}$ (mag) & log g (cgs)\\
  \hspace{0.35cm} $\pm$ error &  \hspace{0.35cm} $\pm$ error &  \hspace{0.45cm }$\pm$ error & $\pm$ error\\
\noalign{\smallskip}\hline\noalign{\smallskip}
M$_{\rm 1}$ = 2.014 & R$_{\rm 1}$ = 1.733 & M$_{{\rm Bol},{\rm 1}}$ = 1.865 & log g$_{\rm 1}$ = 4.264\\
\hspace{0.35cm} $\pm$ 0.004 & \hspace{0.38cm}$\pm$ 0.006 & \hspace{0.8cm}$\pm$ 0.008 & $\pm$ 0.005 \\
M$_{\rm 2}$ = 0.211 & R$_{\rm 2}$ = 2.19 & M$_{{\rm Bol},{\rm 2}}$ = 3.145 & log g$_{\rm 2}$ = 3.078\\ 
\hspace{0.35cm} $\pm$ 0.001 & \hspace{0.35cm} $\pm$ 0.01 & \hspace{0.75cm} $\pm$ 0.008 & $\pm$ 0.003 \\
\hline\noalign{\smallskip}
  radius (pole) &   radius (point) & radius (side) & radius (back)\\
\noalign{\smallskip}\hline\noalign{\smallskip}
r$_{\rm 1}$ = 0.1634 & r$_{\rm 1}$ = 0.1643 & r$_{\rm 1}$ = 0.1641 & r$_{\rm 1}$ = 0.1642\\ 
r$_{\rm 2}$ = 0.1924 & r$_{\rm 2}$ = 0.2860 & r$_{\rm 2}$ = 0.2000 & r$_{\rm 2}$ = 0.2310\\ 
\hline
\end{tabular}
\end{table}

\section{Discussion and conclusions}\label{sec:con}

\textcolor{black}{We determined new system and stellar parameters for the oscillating Algol-type eclipsing binary AS~Eri based on combined photometry and spectroscopy, namely, two photometric data sets from the MOST and TESS missions, as well as a large series of component radial velocities obtained from high-resolution spectroscopy. The final solution was obtained through a series of minimizations, where a comparison with previous studies was done when possible as an independent verification of our results. Here, we performed the final modelling using the refined orbital period of 2.6641496~d which was needed to remove the phase gap between the two light curves from space. The final model agrees very well with the high-quality data of the TESS light curve. We also showed that this solution does not entirely fit the MOST light curve as some of the parameters such as the effective temperature of comp~B, the inclination (i) and the gravitational potential of comp~A ($\Omega_{1}$) needed a significant correction. Both LC models indicate an almost circular orbit. However, the RV data are better represented with a small non-zero eccentricity, which indicates that the space photometric and ground-based spectroscopic data do not fully agree with each other. Since the uncertainties of the RVs were cautiously estimated, this did not prevent us from finding an optimum solution with all three data sets combined into one minimization run.} {On the other hand, we also remarked that the small eccentricity found by fitting the RVs of comp~B could be a substitute for RV distortions that are typical of Algol systems.} \textcolor{black}{The residuals obtained after subtraction of the best model derived in this work will enable to study the pulsation properties of AS~Eri with a higher accuracy and more details than hitherto possible.} \\

\begin{figure}
\center
  \includegraphics[width=8.8cm,clip=]{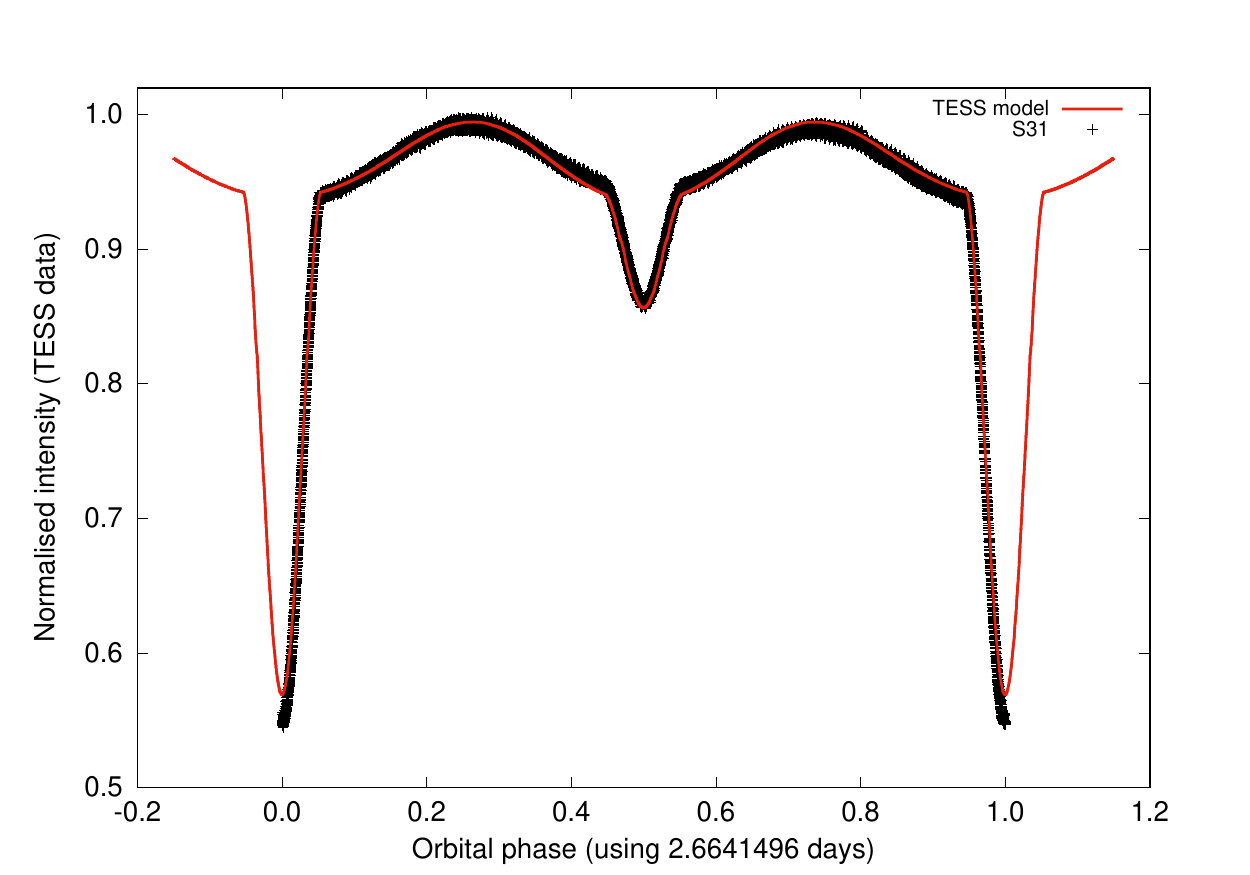} \\
\caption{The folded TESS light curve from Sector 31 overplotted with the TESS model (solid red line).}
\label{fig:ASEri:TESS:S31}
\end{figure}

\textcolor{black}{We conclude with the following remarks:\\ 
- AS~Eri is a system whose mass ratio, q, is very accurately known because both techniques contribute to its determination. Note that the new q is close to that derived by \citet{1984A&A...141....1V}.\\ 
- The final solution agrees overall quite well with the \textcolor{black}{simultaneous} solution (no third light) proposed by VH\&W, except for T$_{{\rm eff},2}$ which is found to be hotter than previously thought, implying a temperature difference of 600 to 800~K between the components. This discrepancy is not due to the release of the gravity brightening parameter, gb$_{2}$, which has profoundly changed from the default value (0.32) to a plausible 0.08 (VH\&W derived a similar low value, see also the case of KIC~9851944 \citep{2016ApJ...826...69G}). The fact that they obtained a different T$_{{\rm eff},2}$ may look surprising, it might be due to using a starting value based on the older findings \citep{1960AJ.....65..139K,1971ApJ...166..373H}.\\
- The final solution corresponds to the semi-detached case where the secondary fills its Roche lobe. The primary component belongs to the main sequence whereas the secondary is an evolved giant star. The relative stellar radii are not compatible with those of VH\&W either, as the mean radius of comp~A is about 10\% larger than theirs. {Our solution is however compatible with the observed RML effect on the condition of asynchronous rotation for comp~A (i.e. with f$_{1}$ = 1.1-1.2). Since the modelled RVs are very sensitive to the synchronicity parameter, we consider the derived value greater than unity as reliable. From this study, we deduce that the orbit is circular (cf. LC models in Sect.~\ref{sec:mod}) and that comp~A is rotating supersynchronously. Supersynchronous rotation of the mass-gaining component has been detected in several Algols \citep[e.g.][]{2010MNRAS.406.1071D}. 
Even a change between super- and subsynchronous rotation has been evidenced during the long-term monitoring of RZ~Cas \citep{2020A&A...644A.121L}. 
It is assumed that the primary or its outer layers are accelerated by the impact of the gas stream from the donor star. This is consistent with a Roche-lobe filling system in the post-rapid phase of mass transfer, for which the (rotational) angular momentum of the components is affected by a past mass transfer event.} \\ 
\textcolor{black}{- The cyclic modulations of the orbital periods observed in a majority of Algol-type systems with late-type, Roche-lobe filling components can be explained by the magnetic activity of their cool companions \citep{1992ApJ...385..621A}. In the case of AS~Eri, we report the high stability of its orbital period based on the absence of any short-term cyclic variation in the (O-C) diagram, which might indicate weak magnetic activity of the secondary component.} 
\textcolor{black}{On the other hand, we showed that the shape of the light curve significantly varies over a time scale of a few years (Fig.~\ref{fig:ASEri:TESS:S31}). 
Therefore, while the orbital period appears to be stable on the long term, we conclude that the light curve is affected by a years-long modulation which is most probably reflecting the magnetic activity cycle of the cool companion. The fact that the orbital period is stable does not contradict this result. }\\ 
- \textcolor{black}{Although the RML effect is well evidenced in the RV residuals of both components (see bottom panel in Fig~\ref{fig:ASEri:orbit0} and also Fig.~\ref{fig:ASEri:wTESS:resRV3}), we are currently unable to detect any asymmetry in the residuals of the primary component. Such asymmetry may be caused by an asymmetrical attenuation of the stellar disk by the gas stream and/or an inhomogeneous distribution of the accreted gas around the primary component \citep{2020A&A...644A.121L}. The observed extremal residuals for the primary component are -7.2 $\pm$ 0.6~\kms~and +5.8 $\pm$ 0.5~\kms. However, a crucial observation is missing at the expected phase of maximum positive deviation (at $\sim$ 0.98). Therefore, we cannot (yet) conclude on the detection of any asymmetric attenuation as an indication of ongoing mass transfer .} \\ 
- An additional finding regarding AS~Eri is the fact that no third light is needed (unlike the alternative simultaneous solution of VH\&W {or the light-curve solution obtained by \citet{2020arXiv200101292A}}). } \\

\section*{Acknowledgements}
This study is based on data collected by the {\it Transiting Exoplanet Survey Satellite} (TESS) mission, described by Jenkins et al.\,(2016) and publicly available from the Mikulski Archive for Space Telescopes (MAST), as well as on data from the {\it Microvariability and Oscillations of STars} (MOST) satellite, a Canadian Space Agency mission, jointly operated by Microsatellite Systems Canada Inc. (MSCI), formerly part of Dynacon, Inc., the University of Toronto Institute for Aerospace Studies, and the University of British Columbia with the assistance of the University of Vienna. Funding for the TESS mission is provided by NASA's Science Mission directorate. This study made use of high-resolution spectra obtained with the HERMES \'echelle spectrograph installed at the {\it Mercator} telescope operated by the IvS, KULeuven, funded by the Flemish Community, and located at the Observatorio Roque de los Muchachos on the island of La Palma, Spain, as well as with the TCES \'echelle spectrograph of the {\it Alfred Jensch} telescope located at the Th\"uringer Landessternwarte, Tautenburg, Germany. This research also made use of the Lichtenknecker-Database of the BAV, operated by the {\it Bundesdeutsche Arbeitsgemeinschaft f\"ur Ver\"anderliche Sterne e.V. (BAV)}. \textcolor{black}{PL gratefully acknowledges the financial support of the Royal Observatory of Belgium to the HERMES Consortium, as well as the help of the HERMES Consortium observers P.~De~Cat, N.~Dries, M.~Hillen, A.~Jorissen and S.~Goriely. DM and KG thankfully acknowledge the support of the National Astronomical Research Institute of Thailand. {HL is grateful for support from the DFG grant with reference number LE 1102/3-1. We thank the referee for most helpful comments.} }

\section*{Data Availability}

\textcolor{black}{The light curve of AS~Eri obtained by the MOST satellite is no longer publicly available. The original data set will be deposited at the {\it Centre de Donn\'ees Stellaires}, Strasbourg, France.}
The data collected by the TESS mission are publicly available from the Mikulski Archive for Space Telescopes (MAST) at \textit{https://mast.stsci.edu/portal/Mashup/Clients/Mast/Portal.html}. \\
The high-resolution spectra acquired with the spectrograph \textsc{HERMES}
are available from the HERMES Consortium lead by the Instituut voor Sterrenkunde, University of Leuven, Leuven, Belgium upon specific request. Similarly, the spectra acquired with the \textsc{TCES} spectrograph 
are available from the Th\"uringer Landessternwarte, Tautenburg, Germany, upon specific request. \\




\bibliographystyle{mnras}
\bibliography{ASEri_bibliography} 

\appendix
\section{The (O-C) data}

\begin{table*}
\caption{Times of minima and their corresponding O-C residuals. {The full table is available online as Supplementary Material.} } 
\begin{tabular}{@{}lcrrcccc@{}}
\hline\noalign{\smallskip}
Time of min. [HJD] & Type  &  Cycle  & (O-C) [d] & Meth. &  Observer & Source & Error \\ \noalign{\smallskip}\hline\noalign{\smallskip}
2415957.885 & P & -15246   & -0.07129 & pg & S.Gaposchkin & HB 918.13 &  ---  \\
2416775.814 & P & -14939   & -0.03639 & pg & S.Gaposchkin & HB 918.13 &  ---  \\
2417140.855 & P & -14802   &  0.01603 & pg & S.Gaposchkin & HB 918.13 &  ---  \\
2417521.778 & P & -14659   & -0.03444 & pg & S.Gaposchkin & HB 918.13 &  ---  \\
2418560.825 & P & -14269   & -0.00601 & pg & S.Gaposchkin & HB 918.13 &  ---  \\
2418592.796 & P & -14257   & -0.00481 & pg & S.Gaposchkin & HB 918.13 &  ---  \\
2418600.785 & P & -14254   & -0.00826 & pg & S.Gaposchkin & HB 918.13 &  ---  \\
2420417.794 & P & -13572   &  0.05031 & pg & S.Gaposchkin & HB 918.13 &  ---  \\
2421576.612 & P & -13137   & -0.03701 & pg & S.Gaposchkin & HB 918.13 &  ---  \\
2423004.599 & P & -12601   & -0.03451 & pg & S.Gaposchkin & HB 918.13 &  ---  \\
\ldots \\
\noalign{\smallskip}\hline\noalign{\smallskip}
\end{tabular}
\label{tab:OminC2}
\end{table*}



\bsp	
\label{lastpage}
\end{document}